\documentclass[conference]{IEEEtran}
\IEEEoverridecommandlockouts

\usepackage{cite}
\usepackage{amsmath,amssymb,amsfonts}
\usepackage{graphicx}
\usepackage{textcomp}
\usepackage{xcolor}
\usepackage{algorithm}
\usepackage{algorithmic}
\usepackage{booktabs}
\usepackage{multirow}
\usepackage{subcaption}

\def\BibTeX{{\rm B\kern-.05em{\sc i\kern-.025em b}\kern-.08em
    T\kern-.1667em\lower.7ex\hbox{E}\kern-.125emX}}

\begin{document}

\title{Client Selection Strategies for Federated Semantic Communications in Heterogeneous IoT Networks}

\author{\IEEEauthorblockN{Samer Lahoud}
\IEEEauthorblockA{\textit{Faculty of Computer Science} \\
\textit{Dalhousie University}, Halifax, Canada\\
sml@dal.ca
}
\and
\IEEEauthorblockN{Kinda Khawam}
\IEEEauthorblockA{\textit{Laboratoire DAVID} \\
\textit{Université de Versailles }, Versailles, France\\
kinda.khawam@uvsq.fr
}

}

\maketitle

\begin{abstract}
The exponential growth of IoT devices presents critical challenges in bandwidth-constrained wireless networks, particularly regarding efficient data transmission and privacy preservation. This paper presents a novel federated semantic communication (SC) framework that enables collaborative training of bandwidth-efficient models for image reconstruction across heterogeneous IoT devices. By leveraging SC principles to transmit only semantic features, our approach dramatically reduces communication overhead while preserving reconstruction quality. We address the fundamental challenge of client selection in federated learning environments where devices exhibit significant disparities in dataset sizes and data distributions. Our framework implements three distinct client selection strategies that explore different trade-offs between system performance and fairness in resource allocation. The system employs an end-to-end SC architecture with semantic bottlenecks, coupled with a loss-based aggregation mechanism that naturally adapts to client heterogeneity. Experimental evaluation on image data demonstrates that while Utilitarian selection achieves the highest reconstruction quality, Proportional Fairness maintains competitive performance while significantly reducing participation inequality and improving computational efficiency. These results establish that federated SC can successfully balance reconstruction quality, resource efficiency, and fairness in heterogeneous IoT deployments, paving the way for sustainable and privacy-preserving edge intelligence applications.
\end{abstract}


\section{Introduction}
The proliferation of Internet of Things (IoT) devices has led to an unprecedented surge in data generation, imposing significant strain on wireless networks. Traditional communication systems, designed for bit-perfect transmission, prove increasingly inefficient for bandwidth-constrained IoT environments, particularly for data-intensive applications such as image and video streaming. This fundamental mismatch between data generation rates and available bandwidth necessitates a paradigm shift in how we approach wireless communication.

Semantic communication (SC) \cite{9679803} has emerged as a promising solution to this challenge. Unlike conventional approaches that prioritize bit-level accuracy, SC focuses on transmitting only the essential meaning or semantic information of data. By moving beyond Shannon's classical information theory, SC directly addresses the significance of conveyed information rather than its precise bit-level reproduction. This approach can drastically reduce communication overhead while preserving task-relevant content, making it particularly suitable for resource-constrained IoT deployments.

Simultaneously, Federated Learning (FL) \cite{9415623} addresses another critical concern in IoT networks: data privacy. FL enables collaborative training of machine learning models across distributed devices without requiring raw data centralization. This distributed paradigm naturally aligns with IoT architectures where data is inherently generated and stored across numerous edge devices. However, the integration of FL with SC systems introduces unique challenges, particularly in heterogeneous environments where devices vary significantly in dataset sizes and data distributions.

The heterogeneity inherent in IoT networks fundamentally impacts federated training dynamics. Without careful client selection mechanisms, naive strategies lead to biased models, slow convergence, and inequitable resource utilization. Some clients may disproportionately bear the computational burden while others remain underutilized, creating both efficiency and fairness concerns. These challenges are particularly acute when combining FL with SC, as the quality of semantic representations directly depends on the diversity and quality of participating clients' data.

This paper investigates a novel federated SC framework tailored for image reconstruction in heterogeneous IoT networks. The system enables collaborative training of SC models across distributed clients without sharing raw data, addressing both bandwidth constraints and privacy concerns. Our primary contribution lies in implementing and analyzing three distinct client selection strategies: (i) Baseline, ensuring equal participation as a fairness benchmark; (ii) Utilitarian, maximizing performance by prioritizing high-utility clients; and (iii) Proportional Fairness, dynamically balancing participation to achieve equitable resource utilization while maintaining competitive performance. The framework employs loss-based aggregation that adapts to client heterogeneity. Through evaluation on image data, we demonstrate that federated SC can successfully balance reconstruction quality, communication efficiency, and fairness in practical IoT deployments.

\section{Related Works}

Recent advances in SC and FL have shown independent promise for IoT applications, yet their integration with intelligent client selection remains largely unexplored. 

Within the SC context, deep learning-based joint source-channel coding approaches have demonstrated significant bandwidth savings while maintaining semantic fidelity \cite{10328187}. However, these methods traditionally depend on centralized training, requiring devices to share raw, privacy-sensitive data. Conversely, FL enables collaborative training without exposing local data, making it ideal for privacy-preserving IoT applications \cite{9415623}. Yet standard FL faces significant challenges from system heterogeneity. The canonical FedAvg algorithm aggregates client models without considering differences in data quality, computational power, or communication capabilities. While advanced methods like FedProx \cite{fedprox} and MOON \cite{moon} improve convergence in heterogeneous settings, they fail to address communication efficiency or fine-grained fairness requirements critical for diverse IoT ecosystems.

To manage such heterogeneity, client selection has emerged as crucial for practical FL deployments. Recent work \cite{NEURIPS2024_7886b9ba} proposed heterogeneity-guided sampling through gradient analysis, while \cite{fi17010018} achieved improved convergence via multi-tier client organization. However, these approaches either require complex optimization unsuitable for resource-limited IoT devices or focus purely on convergence without considering communication efficiency gains possible through semantic representations. Despite growing interest in SC-FL integration, existing work overlooks intelligent client selection. The authors in \cite{10531097} proposed an efficient framework for training semantic communication systems in federated settings, while \cite{10559618} introduced federated contrastive learning for personalized semantic communication. However, these efforts primarily focus on system design and training efficiency, neglecting the fundamental challenge of intelligent client selection that leverages semantic characteristics.

Current approaches miss a critical opportunity: client selection strategies that jointly consider statistical heterogeneity and semantic representation quality. To our knowledge, we present the first comprehensive framework that integrates SC and FL while explicitly addressing client heterogeneity through adaptive selection strategies. Our contributions are threefold: i) We propose a federated semantic communication framework for IoT that achieves significant bandwidth reduction while maintaining data privacy through distributed training. ii) We introduce a loss-based aggregation mechanism adapted to client heterogeneity. iii) We design and evaluate client selection strategies that explore different trade-offs between system performance and fairness, extending beyond existing work to provide practical solutions for heterogeneous IoT deployments.

\section{System Model}

We present a federated SC framework for bandwidth-constrained IoT networks consisting of $K$ distributed clients and a central aggregation server. Each client $k \in \{1, 2, \ldots, K\}$ maintains a private image dataset $\mathcal{D}_k$ and participates in collaborative training without sharing raw data. The global dataset $\mathcal{D} = \bigcup_{k=1}^K \mathcal{D}_k$ is distributed across clients using a Dirichlet distribution with concentration parameter $\alpha > 0$ to simulate non-IID conditions typical of real-world IoT deployments. Each local dataset $\mathcal{D}_k = \{\mathbf{x}_k^{(i)}\}_{i=1}^{|\mathcal{D}_k|}$ contains images $\mathbf{x}_k^{(i)} \in \mathbb{R}^{C \times H \times W}$, where $C$ denotes the number of channels ({\it e.g.}, $C=3$ for RGB) and $H \times W$ represents the spatial dimensions. The heterogeneous data distribution ensures that clients possess varying quantities of images with different statistical properties, reflecting practical scenarios where IoT devices capture data under diverse conditions.

\subsection{Semantic Communication Architecture}

Each client implements an end-to-end SC system parameterized by $\boldsymbol{\Theta} = \{\boldsymbol{\theta}_{\text{se}}, \boldsymbol{\theta}_{\text{ce}}, \boldsymbol{\theta}_{\text{cd}}, \boldsymbol{\theta}_{\text{sd}}\}$, comprising semantic encoder, channel encoder, channel decoder, and semantic decoder parameters. The SC system operates through five stages as detailed below:
 
\paragraph{Semantic Encoding} The encoder $f_{\text{se}}(\cdot; \boldsymbol{\theta}_{\text{se}})$ extracts task-relevant features while preserving reconstruction-critical information:
\begin{equation}
\mathbf{z}, \mathcal{S} = f_{\text{se}}(\mathbf{x}; \boldsymbol{\theta}_{\text{se}}),
\end{equation}
where $\mathbf{z} \in \mathbb{R}^{d_s}$ represents the semantic bottleneck with dimension $d_s \ll C \cdot H \cdot W$, providing a compact representation of the input image $\mathbf{x} \in \mathbb{R}^{C \times H \times W}$.  The set $\mathcal{S} = \{\mathbf{s}_1, \mathbf{s}_2\}$ denotes intermediate feature maps preserved for skip connections.

\paragraph{Channel Encoding} The encoder $f_{\text{ce}}(\cdot; \boldsymbol{\theta}_{\text{ce}})$ further compresses semantic features:
\begin{equation}
\mathbf{x}_c = f_{\text{ce}}(\mathbf{z}; \boldsymbol{\theta}_{\text{ce}}),
\end{equation}
where $\mathbf{x}_c \in \mathbb{R}^{d_c}$ represents the transmitted symbols with $d_c < d_s$, achieving compression ratio $\rho = d_s/d_c$.

\paragraph{Wireless Transmission} The channel introduces Rayleigh fading and additive white Gaussian noise (AWGN):
\begin{equation}
\mathbf{y}_c = \mathbf{h} \odot \mathbf{x}_c + \mathbf{n},
\end{equation}
where $\mathbf{h} \in \mathbb{C}^{d_c}$ contains independent complex Gaussian fading coefficients with $h_i \sim \mathcal{CN}(0, 1)$, $\mathbf{n} \sim \mathcal{CN}(\mathbf{0}, \sigma^2\mathbf{I})$ represents AWGN, and $\odot$ denotes element-wise multiplication. The noise variance is $\sigma^2 = 10^{-\gamma/10}$ for signal-to-noise ratio (SNR) $\gamma$ in dB.

\paragraph{Channel Decoding} Zero-forcing equalization recovers the transmitted symbols:
\begin{equation}
\hat{\mathbf{x}}_c = \mathbf{y}_c \oslash (\mathbf{h} + \epsilon),
\end{equation}
where $\oslash$ denotes element-wise division and $\epsilon$ ensures numerical stability. The channel decoder reconstructs semantic features:
\begin{equation}
\hat{\mathbf{z}} = f_{\text{cd}}(\hat{\mathbf{x}}_c; \boldsymbol{\theta}_{\text{cd}}).
\end{equation}

\paragraph{Semantic Decoding}
Finally, the semantic decoder combines recovered features with skip connections  to reconstruct the image:
\begin{equation}
\hat{\mathbf{x}} = f_{\text{sd}}(\hat{\mathbf{z}}, \mathcal{S}; \boldsymbol{\theta}_{\text{sd}}).
\end{equation}

\subsection{Federated Learning Protocol}
\label{sec:fl-protocol}

\subsubsection{Overview}
The framework operates over $T$ rounds, with the server orchestrating collaborative training through adaptive client selection. Algorithm~\ref{alg:fed_semcom} presents the complete protocol. In each round $t$, the server first selects a subset $\mathcal{S}_t \subseteq \{1, \ldots, K\}$ of clients and assigns each selected client $k$ a specific number of local training epochs $E_k^{(t)}$ based on the chosen strategy (detailed in Section~\ref{sec:client-select}). The server then broadcasts the current global model $\boldsymbol{\Theta}^{(t-1)}$ to all selected clients.

\begin{algorithm}[ht]
\caption{Federated SC with Client Selection}
\label{alg:fed_semcom}
\begin{algorithmic}[1]
\STATE Initialize global model $\boldsymbol{\Theta}^{(0)}$, participation counts $\mathbf{n} = \mathbf{0}$
\FOR{round $t = 1, \ldots, T$}
\STATE Select subset $\mathcal{S}_t \subseteq \{1, \ldots, K\}$ and epochs $\{E_k^{(t)}\}_{k \in \mathcal{S}_t}$ via client selection strategy
\STATE Broadcast $\boldsymbol{\Theta}^{(t-1)}$ to all clients in $\mathcal{S}_t$
\FOR{each client $k \in \mathcal{S}_t$ \textbf{in parallel}}
\STATE $\boldsymbol{\Theta}_k^{(t)} \leftarrow$ \textsc{LocalTrain}($\boldsymbol{\Theta}^{(t-1)}$, $\mathcal{D}_k$, $E_k^{(t)}$)
\STATE $\mathcal{L}_k^{(t)} \leftarrow$ average loss on $\mathcal{D}_k$ using $\boldsymbol{\Theta}_k^{(t)}$
\ENDFOR
\STATE Update participation: $n_k \leftarrow n_k + 1$ for all $k \in \mathcal{S}_t$
\STATE $\boldsymbol{\Theta}^{(t)} \leftarrow$ \textsc{FedAggregate}($\{\boldsymbol{\Theta}_k^{(t)}\}_{k \in \mathcal{S}_t}$, $\{\mathcal{L}_k^{(t)}\}_{k \in \mathcal{S}_t}$)
\ENDFOR
\RETURN Final model $\boldsymbol{\Theta}^{(T)}$
\end{algorithmic}
\end{algorithm}

Each client $k \in \mathcal{S}_t$ performs local optimization on its private dataset $\mathcal{D}_k$ (detailed in Section~\ref{sec:local-optim}), producing updated parameters $\boldsymbol{\Theta}_k^{(t)}$ and reporting its average training loss $\mathcal{L}_k^{(t)}$. After local training completes, the server aggregates client contributions using the \textsc{FedAggregate} procedure shown in Algorithm~\ref{alg:fedaggregate}. 

Our aggregation mechanism employs loss-based weighting to naturally adapt to client heterogeneity. The weight formula $w_k = \frac{1}{|\mathcal{S}_t| - 1}(1 - \frac{\mathcal{L}_k}{\mathcal{L}_{\text{total}}})$ assigns higher weights to clients with lower training loss, effectively prioritizing models that achieve better local performance. This approach offers several advantages: (i) it automatically accounts for data quality differences without requiring explicit data statistics, (ii) it adapts dynamically as client performance evolves during training, and (iii) it requires no additional hyperparameter tuning. Unlike standard FedAvg which treats all clients equally, our loss-based aggregation naturally handles non-IID data distributions by giving more influence to clients whose local data leads to better model convergence.

\begin{algorithm}[ht]
\caption{\textsc{FedAggregate} Procedure}
\label{alg:fedaggregate}
\begin{algorithmic}[1]
\REQUIRE Client models $\{\boldsymbol{\Theta}_k\}_{k \in \mathcal{S}_t}$, losses $\{\mathcal{L}_k\}_{k \in \mathcal{S}_t}$
\STATE Compute total loss: $\mathcal{L}_{\text{total}} \leftarrow \sum_{k \in \mathcal{S}_t} \mathcal{L}_k$
\FOR{each client $k \in \mathcal{S}_t$}
\STATE $w_k \leftarrow \frac{1}{|\mathcal{S}_t| - 1} \left(1 - \frac{\mathcal{L}_k}{\mathcal{L}_{\text{total}} + \epsilon}\right)$, where $\epsilon = 10^{-8}$
\ENDFOR
\STATE $\boldsymbol{\Theta} \leftarrow \sum_{k \in \mathcal{S}_t} w_k \cdot \boldsymbol{\Theta}_k$
\RETURN Aggregated model $\boldsymbol{\Theta}$
\end{algorithmic}
\end{algorithm}

\subsubsection{Local Optimization}
\label{sec:local-optim}

During round $t$, each selected client $k \in \mathcal{S}_t$ performs $E_k^{(t)}$ epochs of local optimization to minimize the following objective:
\begin{equation}
\min_{\boldsymbol{\Theta}} \frac{1}{|\mathcal{D}_k|} \sum_{\mathbf{x} \in \mathcal{D}_k} \mathcal{L}(\mathbf{x}, \hat{\mathbf{x}}) + \mu \|\boldsymbol{\Theta}\|_2^2,
\end{equation}
where $\hat{\mathbf{x}} = f_{\text{SC}}(\mathbf{x}; \boldsymbol{\Theta})$ represents the reconstructed image through the complete SC system (encoding, transmission, and decoding), and $\mu$ is the weight decay coefficient for regularization. The reconstruction loss $\mathcal{L}$ combines two complementary objectives:
\begin{equation}
\label{eq:reconst-loss}
\mathcal{L}(\mathbf{x}, \hat{\mathbf{x}}) = \alpha \|\mathbf{x} - \hat{\mathbf{x}}\|_2^2 + (1 - \alpha) \|\mathbf{x} - \hat{\mathbf{x}}\|_1,
\end{equation}
where $\alpha \in [0,1]$ balances between $\ell_2$ norm for smooth reconstruction and $\ell_1$ norm for robustness to outliers and preservation of sharp edges.

Training proceeds via gradient-based optimization with mini-batches of size $B$. To ensure stable convergence in the presence of noisy gradients from channel effects, we apply gradient clipping:
\begin{equation}
\nabla \boldsymbol{\Theta} \leftarrow \min\left(1, \frac{\tau}{\|\nabla \boldsymbol{\Theta}\|_2}\right) \nabla \boldsymbol{\Theta},
\end{equation}
where $\tau$ is the clipping threshold. After completing $E_k^{(t)}$ epochs, client $k$ transmits its updated model parameters $\boldsymbol{\Theta}_k^{(t)}$ and average training loss $\mathcal{L}_k^{(t)}$ back to the server for aggregation.

\subsection{Client Selection Strategies}
\label{sec:client-select}

We formulate client selection as an integer linear programming (ILP) problem that allocates a fixed epoch budget across heterogeneous clients. Given $K$ clients and total epoch budget $E_{\text{total}}$ for round $t$, we determine the number of epochs $E_k^{(t)} \in \{0, 1, \ldots, E_{\max}\}$ assigned to each client $k$ and binary participation indicators $x_k^{(t)} \in \{0, 1\}$.

The objective function balances performance maximization with fairness considerations:
\begin{equation}
\max \sum_{k=1}^{K} U_k^{(t)} \cdot E_k^{(t)} - \lambda \sum_{k=1}^{K} n_k^{(t-1)} \cdot x_k^{(t)}
\end{equation}
The first term maximizes total utility-weighted epochs to optimize system performance, while the second term penalizes repeated selection of the same clients to promote fairness. The performance utility combines dataset size and training quality:
\begin{equation}
U_k^{(t)} = \frac{|\mathcal{D}_k|}{\mathcal{L}_k^{(t-1)} + \epsilon}
\end{equation}
This utility function captures both data quantity (larger datasets) and quality (lower loss), giving preference to clients that can contribute high-quality updates. Here, $\mathcal{L}_k^{(t-1)}$ represents client $k$'s average training loss from the previous round, $n_k^{(t-1)}$ tracks cumulative participation count, $\lambda \geq 0$ controls the fairness-efficiency trade-off, and $\epsilon > 0$ ensures numerical stability. For the initial round ($t=1$), we set $\mathcal{L}_k^{(0)} = \mathcal{L}_{\text{init}}$ for all clients, where $\mathcal{L}_{\text{init}}$ is a predefined constant.

The optimization is subject to constraints that ensure feasible allocation:
\begin{align}
\sum_{k=1}^{K} E_k^{(t)} &= E_{\text{total}} \label{eq:budget} \\
E_k^{(t)} &\leq E_{\max} \cdot x_k^{(t)} \quad \forall k \label{eq:link1} \\
E_k^{(t)} &\geq x_k^{(t)} \quad \forall k \label{eq:link2}
\end{align}
The budget constraint \eqref{eq:budget} ensures total epoch allocation matches available resources. Linking constraints \eqref{eq:link1}-\eqref{eq:link2} couple participation decisions with epoch assignments, ensuring that $x_k^{(t)} = 1$ if and only if client $k$ is selected (i.e., $E_k^{(t)} > 0$). 

This unified framework encompasses three distinct strategies through parameter selection:
\begin{itemize}
\item \emph{Baseline}: All clients participate equally with $E_k^{(t)} = E_{\text{total}}/K$ and $x_k^{(t)} = 1$ for all $k$, serving as a fairness benchmark.
\item \emph{Utilitarian}: Setting $\lambda = 0$ maximizes total performance by favoring clients with high utility scores, potentially at the cost of participation equity.
\item \emph{Proportional Fairness}: Setting $\lambda > 0$ balances performance with equitable participation by penalizing frequently selected clients, promoting long-term fairness.
\end{itemize}

The ILP formulation maintains computational tractability with $2K$ variables and $2K+1$ constraints, enabling efficient solution using standard solvers for typical FL scenarios with up to 100 clients. After solving the ILP, selected clients $\mathcal{S}_t = \{k : E_k^{(t)} > 0\}$ participate in the federated round with their optimal epoch assignments. The server updates participation history as $n_k^{(t)} = n_k^{(t-1)} + x_k^{(t)}$, maintaining a record for future fairness considerations. This approach provides globally optimal epoch allocation while offering principled control over the fairness-efficiency trade-off through the $\lambda$ parameter.

\section{Results and Discussion}
\label{sec:results}

\subsection{Experimental Setup}

We evaluate our framework using RGB images of dimension $64 \times 64 \times 3$ from the Tiny ImageNet dataset \cite{russakovsky2015imagenet}. The dataset contains 20 classes partitioned across $K = 10$ clients using a Dirichlet distribution with concentration parameter $\alpha = 1.0$ to simulate non-IID data distributions.

The semantic encoder consists of three convolutional layers with channels (3 → 32 → 64 → 128) using stride-2 downsampling and ReLU activations, reducing spatial dimensions from $64 \times 64$ to $8 \times 8$. A fully-connected layer then maps the flattened features to the semantic bottleneck $\mathbf{z} \in \mathbb{R}^{256}$. The channel encoder uses three fully-connected layers (256 → 128 → 64 → 32) to compress features to $d_c = 32$ symbols, achieving an 8× compression ratio. The channel decoder mirrors this with layers (32 → 64 → 128 → 256), while the semantic decoder employs transposed convolutions with skip connections from the first two encoder layers to preserve spatial details.

Each federated round allocates $E_{\text{total}} = 30$ epochs across selected clients. Local optimization employs Adam optimizer with learning rate $\eta = 3 \times 10^{-4}$, weight decay $\mu = 10^{-4}$, batch size $B = 16$, and gradient clipping $\tau = 1.0$. The reconstruction loss uses $\alpha = 0.8$ to emphasize $\ell_2$ reconstruction quality. The wireless channel operates at SNR $\gamma = 10$ dB with Rayleigh fading. We run each experiment for $T = 50$ rounds and report results averaged over 3 random seeds.


\subsection{Performance Metrics}

We employ two sets of metrics to comprehensively evaluate system performance. First, reconstruction quality is assessed using standard metrics computed on a validation set:
\begin{align}
\text{MSE} &= \frac{1}{N \cdot C \cdot H \cdot W} \sum_{i=1}^{N} \|\mathbf{x}_i - \hat{\mathbf{x}}_i\|_2^2 \\
\text{PSNR} &= 10 \log_{10}\left(\frac{1}{\text{MSE}}\right) \text{ dB}
\end{align}
where $N$ is the number of test images and pixel values are normalized to [0,1].

Second, system fairness is quantified through Gini coefficients measuring inequality in client participation and computational effort:
\begin{equation}
\mathcal{G} = \frac{2\sum_{i=1}^{K} i \cdot v_i^{\text{sorted}}}{K \sum_{i=1}^{K} v_i} - \frac{K+1}{K}
\end{equation}
where $v_i^{\text{sorted}}$ represents values sorted in ascending order. We track:
\begin{itemize}
\item \emph{Participation Gini} $\mathcal{G}_{\text{part}}$: Measures inequality in selection frequency across clients
\item \emph{Effort Gini} $\mathcal{G}_{\text{effort}}$: Measures inequality in total computational burden (dataset size × epochs)
\end{itemize}
Both coefficients range from 0 (perfect equality) to 1 (maximum inequality).

\subsection{Numerical Results}

\begin{figure}[t]
  \centering
  \includegraphics[width=0.9\columnwidth]{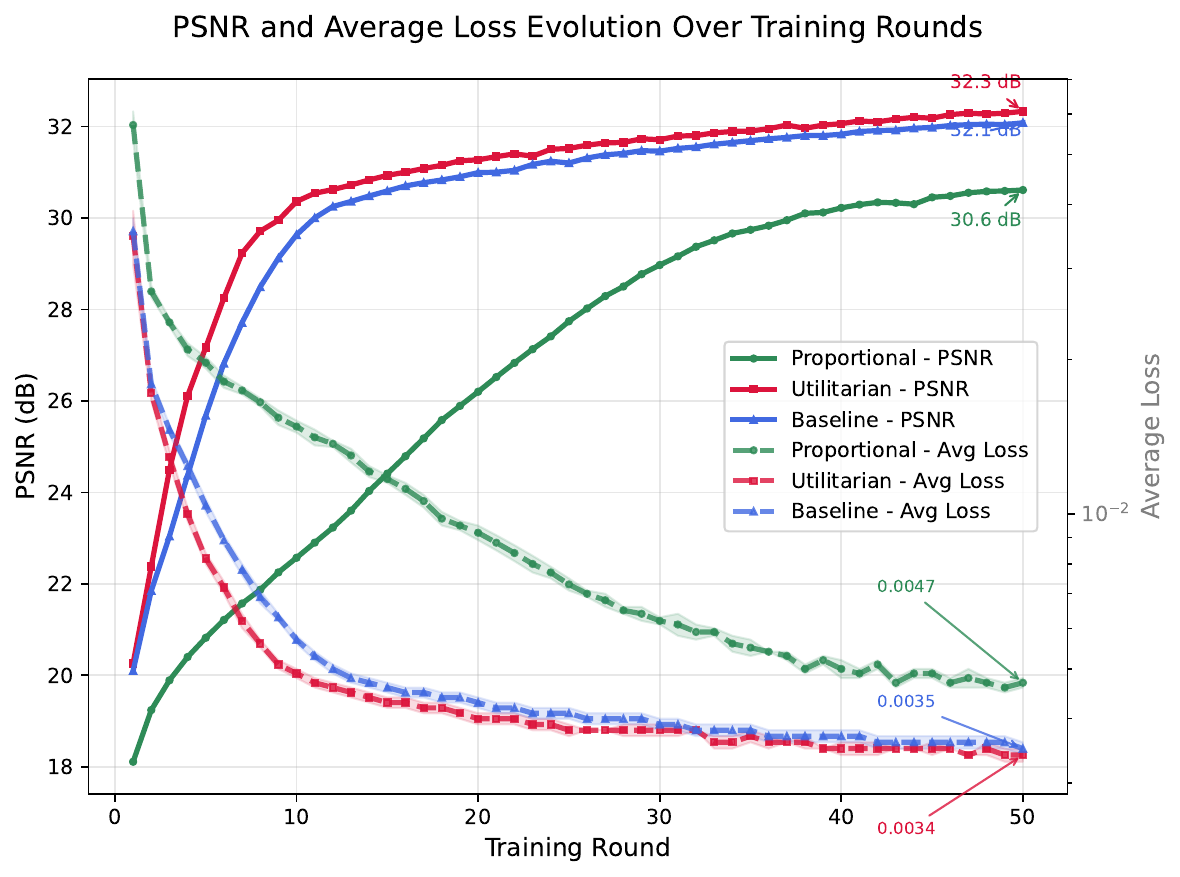}
  \caption{PSNR and average loss evolution over 50 rounds.}
  \label{fig:psnr}
\end{figure}

Figure~\ref{fig:psnr} demonstrates PSNR and average loss evolution over 50 federated rounds. Utilitarian selection achieves the highest terminal PSNR (32.3 dB), marginally outperforming Baseline (32.1 dB), while Proportional Fairness converges to 30.6 dB. Both Utilitarian and Baseline exhibit rapid initial convergence, reaching 30 dB within 10 rounds, whereas Proportional Fairness follows a steadier trajectory reflecting its balanced client selection. The average reconstruction loss curves given in \eqref{eq:reconst-loss} (shown on the secondary axis) corroborate these findings, with Utilitarian and Baseline achieving losses below $4 \times 10^{-3}$ early in training, while Proportional Fairness converges more gradually.

\begin{figure}[t]
  \centering
  \includegraphics[width=0.9\columnwidth]{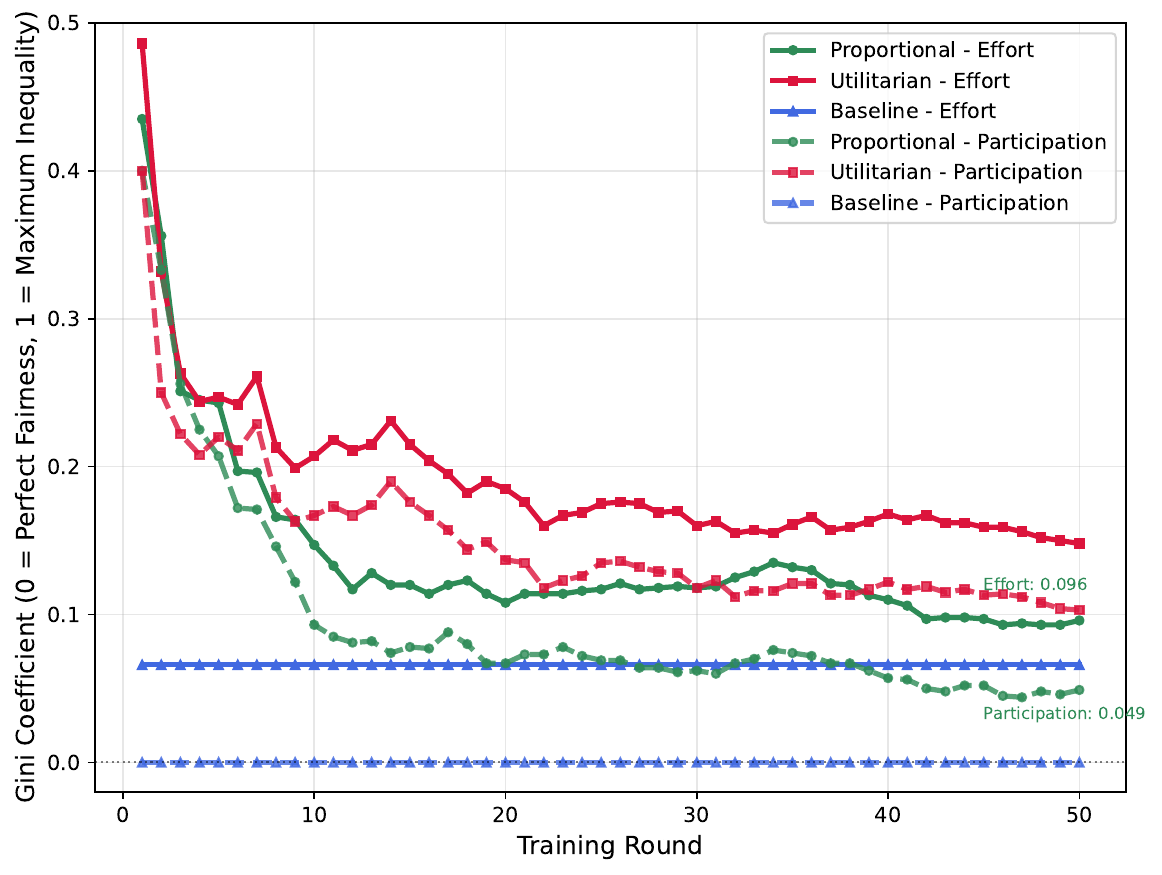}
  \caption{Gini coefficients for effort and participation fairness.}
  \label{fig:gini}
\end{figure}

Figure~\ref{fig:gini} tracks fairness evolution through Gini coefficients. Baseline maintains perfect participation equity ($\mathcal{G}_{\text{part}} = 0$) and minimal effort inequality ($\mathcal{G}_{\text{effort}} < 0.06$) by design. Utilitarian selection exhibits persistent bias, stabilizing at $\mathcal{G}_{\text{effort}} \approx 0.15$ for effort and $\mathcal{G}_{\text{part}} \approx 0.10$ for participation after initial fluctuations. In contrast, Proportional Fairness achieves substantial inequality reduction, converging to $\mathcal{G}_{\text{effort}} = 0.096$ and $\mathcal{G}_{\text{part}} = 0.049$, representing a 36\% improvement in effort equality compared to Utilitarian while maintaining competitive reconstruction quality.

\begin{figure}[t]
  \centering
  \includegraphics[width=0.9\columnwidth]{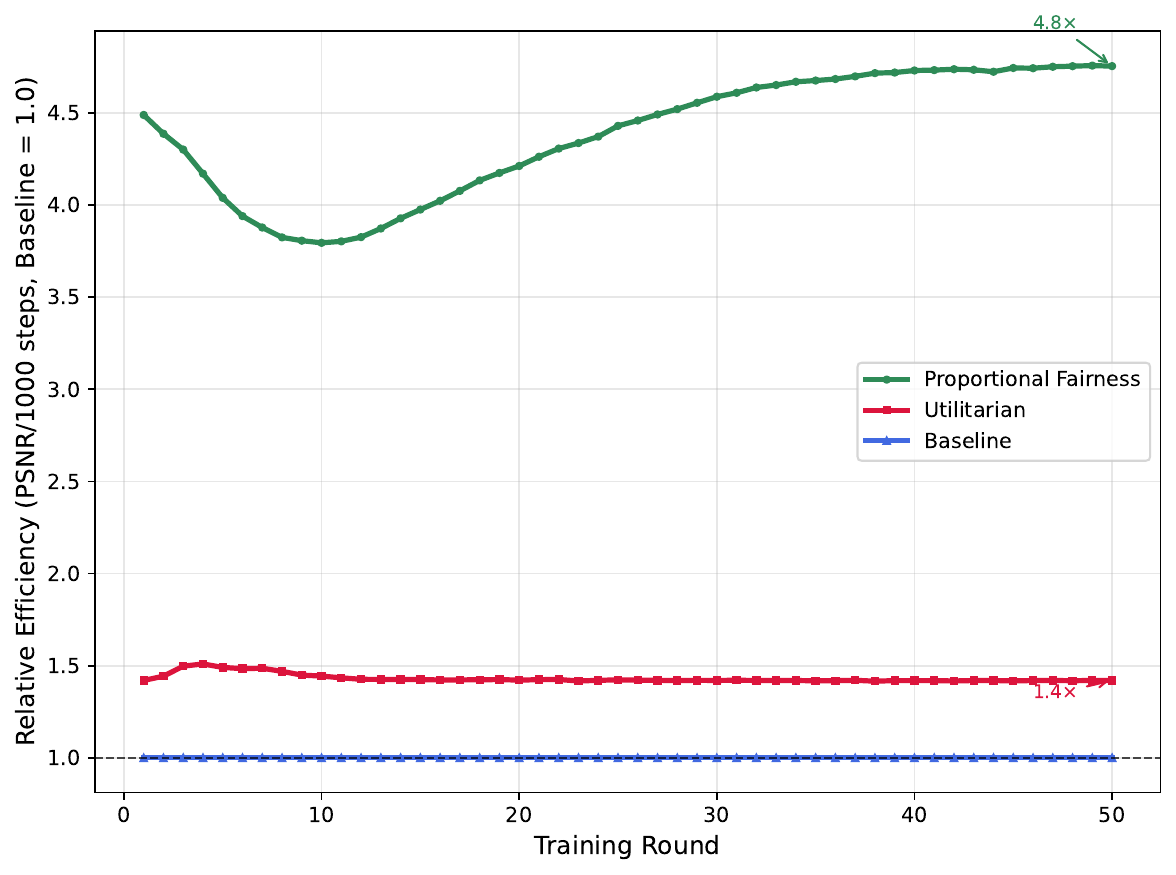}
  \caption{Relative computational efficiency.}
  \label{fig:eff}
\end{figure}

Figure~\ref{fig:eff} presents computational efficiency, measured as PSNR per 1000 training steps (dataset size × epochs), normalized to Baseline performance. This metric captures how effectively each strategy utilizes computational resources to improve model quality. 
Proportional Fairness demonstrates superior resource utilization with 4.8× efficiency gain, substantially exceeding Utilitarian's 1.4× improvement. This advantage becomes pronounced after round 10, where Proportional Fairness shows an initial dip before steadily improving. The divergence suggests that Utilitarian's greedy selection provides limited efficiency gains, while Proportional Fairness's balanced approach yields compounding benefits over time.

\section{Conclusion}
This paper presented a federated semantic communication framework for heterogeneous IoT networks, achieving 8× bandwidth reduction while preserving data privacy. Client selection in such heterogeneous environments fundamentally impacts both system performance and fairness. Our analysis revealed critical trade-offs: Utilitarian selection maximized PSNR (32.3 dB) but exhibited participation bias, while Proportional Fairness achieved competitive performance (30.6 dB) with 36\% reduced inequality and 4.8× better computational efficiency. Future work should explore game-theoretic frameworks for client selection where IoT devices strategically decide participation based on individual utilities. Coalition formation among resource-constrained devices could enable collaborative training while ensuring fair reward distribution. This work establishes federated SC as a practical paradigm for sustainable edge intelligence in bandwidth-constrained environments.

\bibliographystyle{ieeetr}
\bibliography{semcom-fair}

\end{document}